\documentclass[prl,aps,twocolumn,superscriptaddress,showpacs,preprintnumbers,
               amsmath,amssymb,floatfix]{revtex4}

\def\gtwid{\mathrel{\raise.3ex\hbox{$>$\kern-.75em\lower1ex\hbox{$\sim$}}}}
\def\ltwid{\mathrel{\raise.3ex\hbox{$<$\kern-.75em\lower1ex\hbox{$\sim$}}}}

\usepackage{graphicx}
\usepackage{epstopdf}
\usepackage{dcolumn}
\usepackage{bm}

\def\gev{GeV/c$^2$}

\begin{document}
\title{Results from the Final Exposure of the CDMS~II Experiment}


\affiliation{Division of Physics, Mathematics \& Astronomy, California Institute of Technology, Pasadena, CA 91125, USA} 
\affiliation{Department of Physics, Case Western Reserve University, Cleveland, OH  44106, USA}
\affiliation{Fermi National Accelerator Laboratory, Batavia, IL 60510, USA}
\affiliation{Lawrence Berkeley National Laboratory, Berkeley, CA 94720, USA}
\affiliation{Department of Physics, Massachusetts Institute of Technology, Cambridge, MA 02139, USA}
\affiliation{Department of Physics, Queen's University, Kingston, ON, Canada, K7L 3N6}
\affiliation{Department of Physics, St.\,Olaf College, Northfield, MN 55057 USA}
\affiliation{Department of Physics, Santa Clara University, Santa Clara, CA 95053, USA}
\affiliation{Department of Physics, Southern Methodist University, Dallas, TX 75275, USA}
\affiliation{Department of Physics, Stanford University, Stanford, CA 94305, USA}
\affiliation{Department of Physics, Syracuse University, Syracuse, NY 13244, USA}
\affiliation{Department of Physics, Texas A \& M University, College Station, TX 77843, USA}
\affiliation{Department of Physics, University of California, Berkeley, CA 94720, USA}
\affiliation{Department of Physics, University of California, Santa Barbara, CA 93106, USA}
\affiliation{Departments of Phys. \& Elec. Engr., University of Colorado Denver, Denver, CO 80217, USA}
\affiliation{Department of Physics, University of Florida, Gainesville, FL 32611, USA}
\affiliation{School of Physics \& Astronomy, University of Minnesota, Minneapolis, MN 55455, USA}
\affiliation{Physics Institute, University of Z\"{u}rich, Winterthurerstr. 190, CH-8057, Switzerland}

\author{Z.~Ahmed} \affiliation{Division of Physics, Mathematics, and Astronomy, California Institute of Technology, Pasadena, CA 91125, USA} 
\author{D.S.~Akerib} \affiliation{Department of Physics, Case Western Reserve University, Cleveland, OH  44106, USA} 
\author{S.~Arrenberg} \affiliation{Physics Institute, University of Z\"{u}rich, Winterthurerstr. 190, CH-8057, Switzerland}
\author{C.N.~Bailey} \affiliation{Department of Physics, Case Western Reserve University, Cleveland, OH  44106, USA} 
\author{D.~Balakishiyeva} \affiliation{Department of Physics, University of Florida, Gainesville, FL 32611, USA} 
\author{L.~Baudis} \affiliation{Physics Institute, University of Z\"{u}rich, Winterthurerstr. 190, CH-8057, Switzerland}
\author{D.A.~Bauer} \affiliation{Fermi National Accelerator Laboratory, Batavia, IL 60510, USA} 
\author{P.L.~Brink} \affiliation{Department of Physics, Stanford University, Stanford, CA 94305, USA} 
\author{T.~Bruch} \affiliation{Physics Institute, University of Z\"{u}rich, Winterthurerstr. 190, CH-8057, Switzerland}
\author{R.~Bunker} \affiliation{Department of Physics, University of California, Santa Barbara, CA 93106, USA} 
\author{B.~Cabrera} \affiliation{Department of Physics, Stanford University, Stanford, CA 94305, USA} 
\author{D.O.~Caldwell} \affiliation{Department of Physics, University of California, Santa Barbara, CA 93106, USA} 
\author{J.~Cooley} \affiliation{Department of Physics, Southern Methodist University, Dallas, TX 75275, USA} 
\author{P.~Cushman} \affiliation{School of Physics \& Astronomy, University of Minnesota, Minneapolis, MN 55455, USA} 
\author{M.~Daal} \affiliation{Department of Physics, University of California, Berkeley, CA 94720, USA} 
\author{F.~DeJongh} \affiliation{Fermi National Accelerator Laboratory, Batavia, IL 60510, USA} 
\author{M.R.~Dragowsky} \affiliation{Department of Physics, Case Western Reserve University, Cleveland, OH  44106, USA} 
\author{L.~Duong} \affiliation{School of Physics \& Astronomy, University of Minnesota, Minneapolis, MN 55455, USA} 
\author{S. Fallows}\affiliation{School of Physics \& Astronomy, University of Minnesota, Minneapolis, MN 55455, USA} 
\author{E.~Figueroa-Feliciano} \affiliation{Department of Physics, Massachusetts Institute of Technology, Cambridge, MA 02139, USA} 
\author{J.~Filippini} \affiliation{Division of Physics, Mathematics, and Astronomy, California Institute of Technology, Pasadena, CA 91125, USA} 
\author{M.~Fritts} \affiliation{School of Physics \& Astronomy, University of Minnesota, Minneapolis, MN 55455, USA} 
\author{S.R.~Golwala} \affiliation{Division of Physics, Mathematics, and Astronomy, California Institute of Technology, Pasadena, CA 91125, USA} 
\author{D.R.~Grant} \affiliation{Department of Physics, Case Western Reserve University, Cleveland, OH  44106, USA} 
\author{J.~Hall} \affiliation{Fermi National Accelerator Laboratory, Batavia, IL 60510, USA} 
\author{R.~Hennings-Yeomans} \affiliation{Department of Physics, Case Western Reserve University, Cleveland, OH  44106, USA} 
\author{S.A.~Hertel} \affiliation{Department of Physics, Massachusetts Institute of Technology, Cambridge, MA 02139, USA} 
\author{D.~Holmgren} \affiliation{Fermi National Accelerator Laboratory, Batavia, IL 60510, USA} 
\author{L.~Hsu} \affiliation{Fermi National Accelerator Laboratory, Batavia, IL 60510, USA} 
\author{M.E.~Huber} \affiliation{Departments of Phys. \& Elec. Engr., University of Colorado Denver, Denver, CO 80217, USA}
\author{O.~Kamaev}\affiliation{School of Physics \& Astronomy, University of Minnesota, Minneapolis, MN 55455, USA} 
\author{M.~Kiveni} \affiliation{Department of Physics, Syracuse University, Syracuse, NY 13244, USA} 
\author{M.~Kos} \affiliation{Department of Physics, Syracuse University, Syracuse, NY 13244, USA} 
\author{S.W.~Leman} \affiliation{Department of Physics, Massachusetts Institute of Technology, Cambridge, MA 02139, USA} 
\author{R.~Mahapatra} \affiliation{Department of Physics, Texas A \& M University, College Station, TX 77843, USA} 
\author{V.~Mandic} \affiliation{School of Physics \& Astronomy, University of Minnesota, Minneapolis, MN 55455, USA} 
\author{K.A.~McCarthy} \affiliation{Department of Physics, Massachusetts Institute of Technology, Cambridge, MA 02139, USA} 
\author{N.~Mirabolfathi} \affiliation{Department of Physics, University of California, Berkeley, CA 94720, USA} 
\author{D.~Moore} \affiliation{Division of Physics, Mathematics, and Astronomy, California Institute of Technology, Pasadena, CA 91125, USA} 
\author{H.~Nelson} \affiliation{Department of Physics, University of California, Santa Barbara, CA 93106, USA} 
\author{R.W.~Ogburn}\affiliation{Department of Physics, Stanford University, Stanford, CA 94305, USA} 
\author{A.~Phipps}\affiliation{Department of Physics, University of California, Berkeley, CA 94720, USA} 
\author{M.~Pyle} \affiliation{Department of Physics, Stanford University, Stanford, CA 94305, USA} 
\author{X.~Qiu} \affiliation{School of Physics \& Astronomy, University of Minnesota, Minneapolis, MN 55455, USA} 
\author{E.~Ramberg} \affiliation{Fermi National Accelerator Laboratory, Batavia, IL 60510, USA} 
\author{W.~Rau} \affiliation{Department of Physics, Queen's University, Kingston, ON, Canada, K7L 3N6}
\author{A.~Reisetter} \affiliation{School of Physics \& Astronomy, University of Minnesota, Minneapolis, MN 55455, USA} \affiliation{Department of Physics, St.\,Olaf College, Northfield, MN 55057 USA} 
\author{T.~Saab} \affiliation{Department of Physics, University of Florida, Gainesville, FL 32611, USA}
\author{B.~Sadoulet} \affiliation{Lawrence Berkeley National Laboratory, Berkeley, CA 94720, USA} \affiliation{Department of Physics, University of California, Berkeley, CA 94720, USA}
\author{J.~Sander} \affiliation{Department of Physics, University of California, Santa Barbara, CA 93106, USA} 
\author{R.W.~Schnee} \affiliation{Department of Physics, Syracuse University, Syracuse, NY 13244, USA} 
\author{D.N.~Seitz} \affiliation{Department of Physics, University of California, Berkeley, CA 94720, USA} 
\author{B.~Serfass} \affiliation{Department of Physics, University of California, Berkeley, CA 94720, USA} 
\author{K.M.~Sundqvist} \affiliation{Department of Physics, University of California, Berkeley, CA 94720, USA} 
\author{M.~Tarka}\affiliation{Physics Institute, University of Z\"{u}rich, Winterthurerstr. 190, CH-8057, Switzerland}
\author{P.~Wikus} \affiliation{Department of Physics, Massachusetts Institute of Technology, Cambridge, MA 02139, USA} 
\author{S.~Yellin} \affiliation{Department of Physics, Stanford University, Stanford, CA 94305, USA} \affiliation{Department of Physics, University of California, Santa Barbara, CA 93106, USA}
\author{J.~Yoo} \affiliation{Fermi National Accelerator Laboratory, Batavia, IL 60510, USA} 
\author{B.A.~Young} \affiliation{Department of Physics, Santa Clara University, Santa Clara, CA 95053, USA} 
\author{J.~Zhang}\affiliation{School of Physics \& Astronomy, University of Minnesota, Minneapolis, MN 55455, USA} 

\collaboration{CDMS Collaboration}

\noaffiliation

\begin{abstract}
We report results from a blind analysis of the final data taken with the  Cryogenic Dark Matter Search experiment (CDMS~II) at the 
Soudan Underground Laboratory, Minnesota, USA.  A total raw exposure of 612 kg-days was analyzed for this work. 
We observed two events in the signal region; based on our background estimate,
the probability of observing two or more background events is 23\%.
These data set an upper limit on the Weakly Interacting Massive Particle (WIMP)-nucleon elastic-scattering spin-independent cross-section of $7.0 \times 10^{-44}$ cm$^{2}$ for a WIMP of mass 70\,GeV/c$^{2}$ at the 90\% confidence level.  Combining this result with 
all previous CDMS~II data gives an upper limit on the WIMP-nucleon spin-independent cross-section of $3.8 \times 10^{-44}$ cm$^{2}$ 
for a WIMP of mass 70\,GeV/c$^{2}$.  We also exclude new parameter space in recently proposed inelastic dark matter models.
\end{abstract}


\maketitle


Cosmological observations \cite{Komatsu:2008hk} have led to a concordance model of the 
universe where $\sim$85\% of matter is non-baryonic, non-luminous and non-relativistic at the time of structure formation.  
Weakly Interacting Massive Particles (WIMPs)  \cite{Steigman:1984ac} are a class of candidates for this dark matter which are 
particularly well motivated by proposed extensions to the Standard Model of particle physics and by thermal production 
models for dark matter in the early universe \cite{Lee:1977ua, Weinberg:1982id,Jungman:1995df,Bertone:2004pz}.  
WIMPs, distributed in a halo surrounding our galaxy, would coherently scatter off nuclei in terrestrial 
detectors \cite{Goodman:1984dc, Gaitskell:2004gd,Salucci:2002jg} 
with a mean recoil energy of $\sim$ tens of keV, presently limited by observation to a rate
 of less than 0.1 event kg$^{-1}$day$^{-1}$ \cite{Lewin:1995rx, Jungman:1995df, Bertone:2004pz}.  
 Direct search experiments seek recoil signatures of these
interactions and have achieved the sensitivity to begin testing the most interesting classes of WIMP 
 models \cite{Ahmed:2008eu,Armengaud:2009hc,Aprile:2008rc,Lebedenko:2008gb}.

The Cryogenic Dark Matter Search (CDMS~II) experiment, located at the Soudan Underground Laboratory, 
uses 19 Ge ($\sim$230~g) and 11 Si ($\sim$100~g) particle detectors operated at 
cryogenic temperatures ($<50$~mK) \cite{PhysRevD.72.052009, Ahmed:2008eu}. Each detector is a disk $\sim$10~mm thick and 76~mm in diameter.  
Particle interactions in the detectors deposit energy in the form of phonons and ionization.  
Phonon sensors on the top of each detector are connected to four phonon readout channels to allow 
measurement of the recoil energy and position of an event.
These phonon sensors are also the ground reference for the ionization measurement.
The electric field for the ionization measurement is formed by applying a voltage bias to the bottom detector surface, 
which is segmented into two concentric electrodes. 
Events having an ionization signal in the outer ionization channel of the
detector are excluded, defining an ionization fiducial volume. The detectors are grouped into five towers, 
each tower containing six detectors. 
Detectors are identified by their tower number (T1-T5) and their position within that tower (Z1-Z6).
A direct line of sight between adjacent detectors in a tower allows identification of events scattering between detectors.

The ratio of the ionization to recoil energy (``ionization yield'') provides event-by-event rejection of electron-recoil events to better than 10$^{-4}$ misidentification.  Essentially all of the misidentified electron recoils are ``surface events'' occurring within the first few $\mu$m of the detector surface, which can suffer from sufficiently reduced ionization collection to be misclassified as nuclear recoils.  Due to interactions of phonons in the surface metal layers, these surface events have faster-rising phonon pulses than events occurring within the bulk  of the detectors (``bulk events").  Hence, we use phonon pulse timing parameters to improve rejection of surface events to obtain
an overall (yield and timing) misidentification probability of less than 10$^{-6}$ for electron recoils.
To attenuate external environmental radioactivity and to reject events caused by cosmogenic 
muons, the detectors are surrounded by layers of lead and polyethylene shielding and an active muon veto \cite{PhysRevD.72.052009}.

Data taken during four periods of stable operation between July 2007 and September 2008 
were analyzed for this work.   Between each data period the cryostat was warmed up for maintenance of the 
cryogenic system.  All 30 detectors were used to identify particle interactions, but only Ge detectors were used to 
search for WIMP scatters.  Five Ge detectors were not used for WIMP detection because of poor performance 
or insufficient calibration data; four more detectors were similarly excluded during subsets of the four periods.  We excluded Si detectors in this analysis due to their lower sensitivity to coherent nuclear elastic scattering.

A subset of events were analyzed to monitor detector stability and identify periods of poor detector performance.   
Data quality criteria were developed on a detector-by-detector basis using Kolmogorov-Smirnov tests performed 
on parameter distributions.  Our detectors require regular neutralization \cite{PhysRevD.72.052009}
to maintain full ionization collection.  
We monitor the yield distribution and remove periods with poor ionization collection.  After these data quality 
selections, the total exposure to WIMPs considered for this work was 612 kg-days.

\begin{figure}[t]
\begin{center}
\includegraphics[width=3.25in]{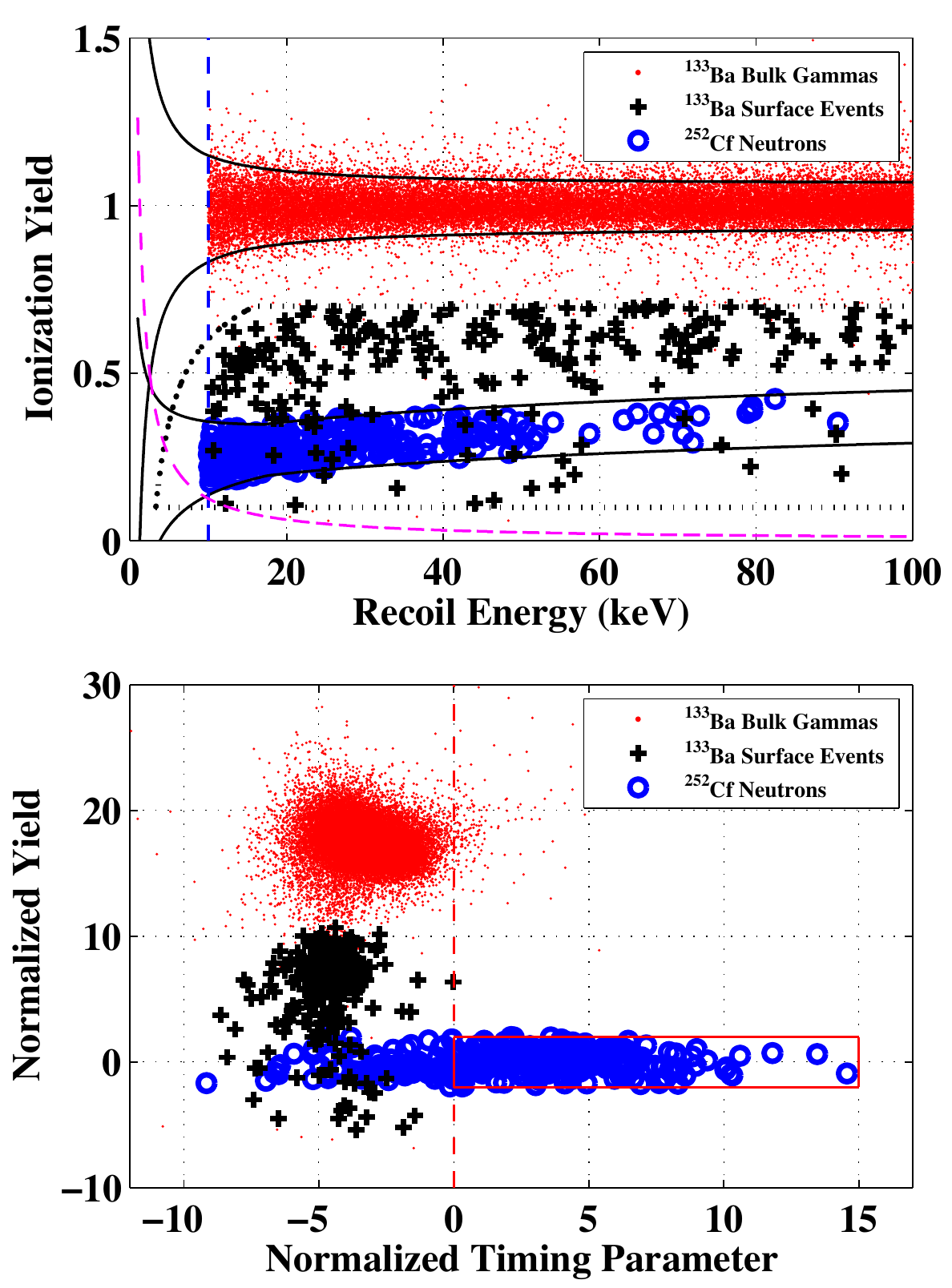}
\caption{The power of the primary background discrimination parameters, ionization yield and phonon timing, is illustrated for a typical 
detector using {\it in situ} calibration sources. 
Shown are bulk electron recoils (red points), surface electron events (black crosses) 
and nuclear recoils (blue circles) with recoil energy between 10 and 100 keV.
Top: Ionization yield versus recoil energy. The solid black lines define bands
that are $2\sigma$ from the mean electron- and nuclear-recoil yields. 
The sloping magenta line indicates the ionization energy threshold while the vertical dashed line 
is the recoil energy analysis threshold. The region enclosed by the black dotted lines defines the sample 
of events that are used to develop surface-event cuts. 
Bottom: Normalized ionization yield (number of standard deviations from mean 
of nuclear recoil band) versus normalized timing parameter (timing relative to 
acceptance region) is shown for the same data.  Events to the right of the 
vertical red dashed line pass the surface-event rejection cut for this detector.
The solid red box is the WIMP signal region. (Color online.)}
\label{fig:calibration}
\end{center}
\end{figure}

The data selection criteria (``cuts") that define the WIMP acceptance region were
 calculated from calibration sets of electron and nuclear recoils obtained during regular
exposures of the detectors to $\gamma$-ray ($^{133}$Ba) and neutron ($^{252}$Cf) sources. 
 A 356 keV $\gamma$-line from the $^{133}$Ba source was used to calibrate the ionization and 
 phonon energy scales of each detector.  Nuclear recoils from the elastic scattering of neutrons emitted by 
 the $^{252}$Cf source were used to define the WIMP signal region, taken to be the 2$\sigma$ band about the 
 mean neutron ionization yield in the yield {\em vs} recoil energy plane (Fig.~\ref{fig:calibration}). 
 Neutron activation of $^{70}$Ge results in the emission of 10.36\,keV x-rays. 
 The width of this x-ray line determined the ionization energy resolution at 
  the 10\,keV analysis threshold to be $\leq 0.4$\,keV on all detectors considered. 

Phonon pulse shapes vary with event position and energy in our detectors.  To enhance surface-event rejection, 
we characterized the variation of the resulting timing and energy estimators using our electron-recoil calibrations 
and corrected for it in both WIMP-search and calibration data. 
To ensure this correction works well at large radius, where position reconstruction degeneracies
can lead to miscorrection, we included calibration electron-recoil events outside the ionization
fiducial volume in the correction calculation. This technique obviated the additional fiducial volume
cut in phonon position reconstruction parameters used in past analyses~\cite{Ahmed:2008eu}.

Figure \ref{fig:calibration} demonstrates our surface-event rejection cut.  
We optimized this cut using calibration sets of nuclear and surface electron
recoils from the $^{252}$Cf and $^{133}$Ba exposures. We verified that the sum of the rise time 
of the largest phonon pulse plus its delay relative to the ionization signal provided the best discrimination between 
surface events and nuclear recoils.   Surface-event rejection criteria based on this discriminator were tuned 
on the calibration data by maximizing the expected sensitivity for a 60\,GeV/c$^2$ 
WIMP.  

A blind analysis was performed, in which cuts were developed without looking at events that  
might appear in or near the signal region. Candidate WIMP-scatters were required to be within 2$\sigma$ of the mean ionization yield of nuclear recoils 
and at least $3\sigma$ away from the mean ionization yield of electron recoils, have recoil energy 
between 10 and 100\,keV, and have ionization energy at least $4.5\sigma$ above the noise.
They are required to occur within the detector
 fiducial volume, satisfy data quality criteria and pass the surface-event rejection cut. Since WIMPs are 
 expected to interact only once (``single-scatter event'') in the experimental apparatus, a candidate event 
 was further required to have energy deposition consistent with noise in 29 detectors. 
 Additionally, we required the absence of significant  
 activity in the surrounding scintillator veto shield during a 200-$\mu s$ window around the trigger.

The efficiency of our analysis cuts for nuclear recoils was measured as a function of energy 
using both neutron-calibration and WIMP-search data.  
The fiducial volume estimate is corrected for neutron multiple-scattering based on simulations.
Our efficiency for signal events has a maximum of $\sim$32\% at 20\,keV. It falls to $\sim$25\% both at 10\,keV, 
due to ionization threshold and flaring of the electron-recoil band; and at 100 keV, due to a drop in fiducial volume. 
The spectrum-averaged equivalent exposure for a WIMP of mass 60\,GeV/c$^{2}$ is 194.1 kg-days.

Neutrons with energies of several MeV can generate single-scatter nuclear recoils that are indistinguishable 
from possible dark matter interactions. Sources of neutron background include cosmic-ray muons interacting 
near the experimental apparatus (outside the veto), radioactive contamination of materials, and 
environmental radioactivity. We performed Monte Carlo simulations of the muon-induced particle showers and subsequent 
neutron production with Geant4~\cite{Agostinelli:2002hh, Allison:2006ve} and 
FLUKA~\cite{Fasso:2003xz, Ferrari:2005zk}.  The cosmogenic background is estimated by multiplying the 
observed number of vetoed single nuclear recoils in the data by the ratio of unvetoed to vetoed 
events as determined by cosmogenic simulation.  This technique predicts $0.04^{+0.04}_{-0.03} $(stat) 
events in this WIMP-search exposure.

Samples of our shielding and detector materials were screened for U and Th
daughters using high purity germanium $\gamma$ counters.  In addition, a global $\gamma$-ray
Monte Carlo was performed and compared to the electromagnetic spectrum
measured by our detectors.  The contamination levels thus determined were
used as input to a Geant4 simulation to calculate the number of neutrons
produced from spontaneous fission and ($\alpha$, n) processes, assuming secular
equilibrium. The estimated background is between 0.03 and 0.06 events and 
is dominated by U spontaneous fission in the copper cans of the cryostat.
The radiogenic neutron background originating from the surrounding rock is 
estimated to be negligibly small compared to other sources.

The number of misidentified surface events was estimated by multiplying 
the observed number of single-scatter events failing the timing cut inside the $2\sigma$ nuclear-recoil band with
the ratio of events expected to pass the timing cut to those failing it (``pass-fail ratio'').
The former was estimated using observed counts from a previous analysis  \cite{Ahmed:2008eu}, and
the latter was estimated using three different methods.
The first method computed the pass-fail ratio from events that reside within the $2\sigma$ nuclear-recoil 
band and multiply scatter in vertically adjacent detectors (``multiple scatter events'').
The second method estimated the pass-fail ratio from multiple-scatter
events surrounding the nuclear-recoil band (``wide-band events'').
Wide-band events have different distributions in energy and in detector face (ionization- or phonon- side) 
from nuclear-recoil band events, affecting the pass-fail ratio. To account for these differences, 
the pass-fail ratio of these events was corrected using the face and energy distributions of events observed in the nuclear-recoil band that failed the timing cut.  
A third, independent estimate of the pass-fail ratio was made using low-yield, multiple-scatter events in $^{133}$Ba 
calibration data, again adjusted for differences in energy and detector-face distributions. 
All three estimates were consistent with each other and were thus combined to obtain an estimate prior to unblinding of
$0.6 \pm 0.1$(stat) surface events misidentified as nuclear recoils.

\begin{figure}[t]
\begin{center}
\includegraphics[width=3.25in]{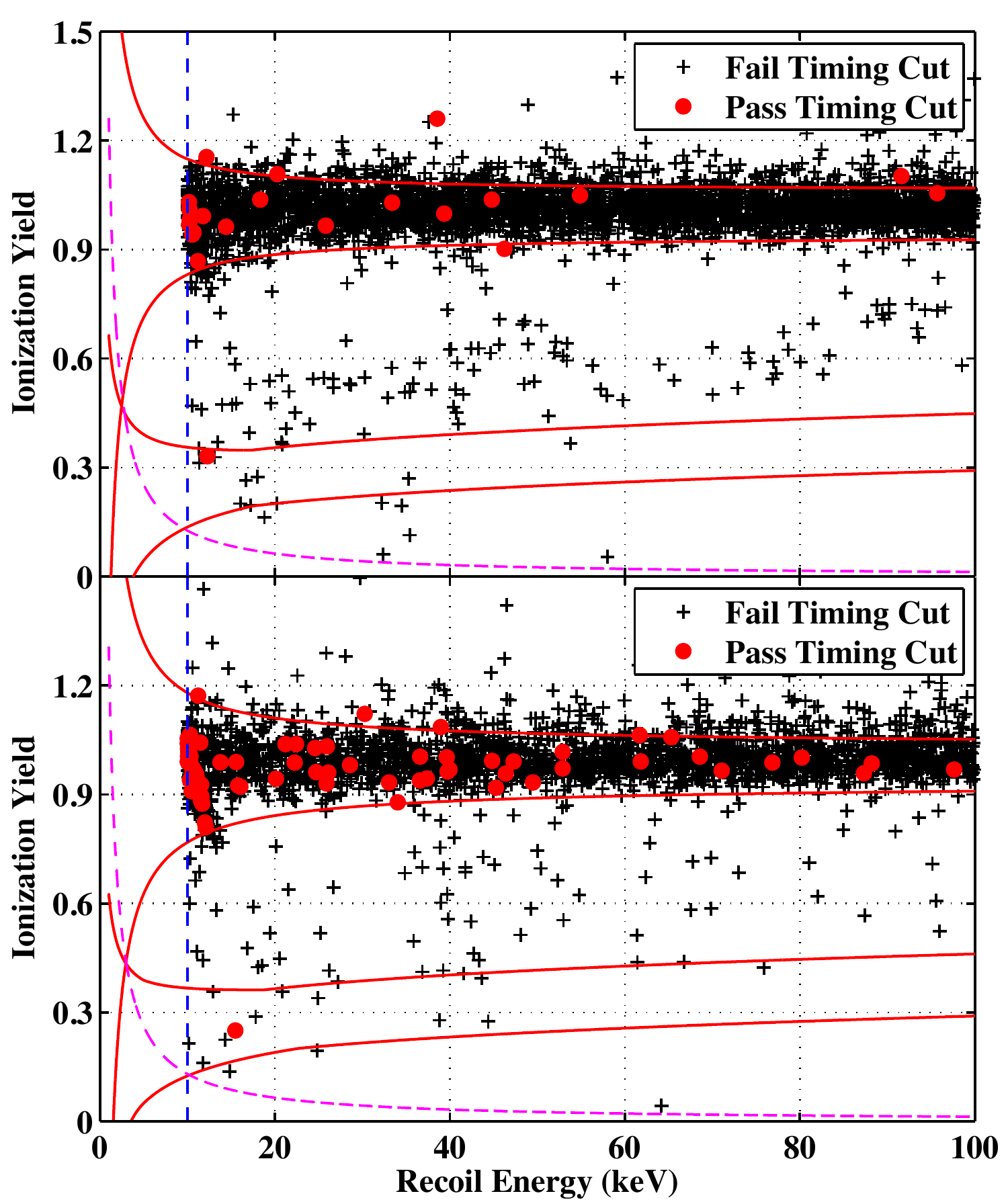}
\caption{Ionization yield versus recoil energy for events passing all cuts, excluding yield and timing.
The top (bottom) plot shows events for detector T1Z5(T3Z4).
The solid red lines indicate the $2\sigma$ electron and nuclear recoil bands. 
The vertical dashed line represents the recoil energy threshold and
the sloping magenta dashed line is the ionization threshold. 
Events that pass the timing cut are shown with round markers.
The candidate events are the round markers inside the nuclear-recoil bands. (Color online.)
}
\label{fig:yield_energy}
\end{center}
\end{figure}

\begin{figure}[t]
\begin{center}
\includegraphics[width=3.25in]{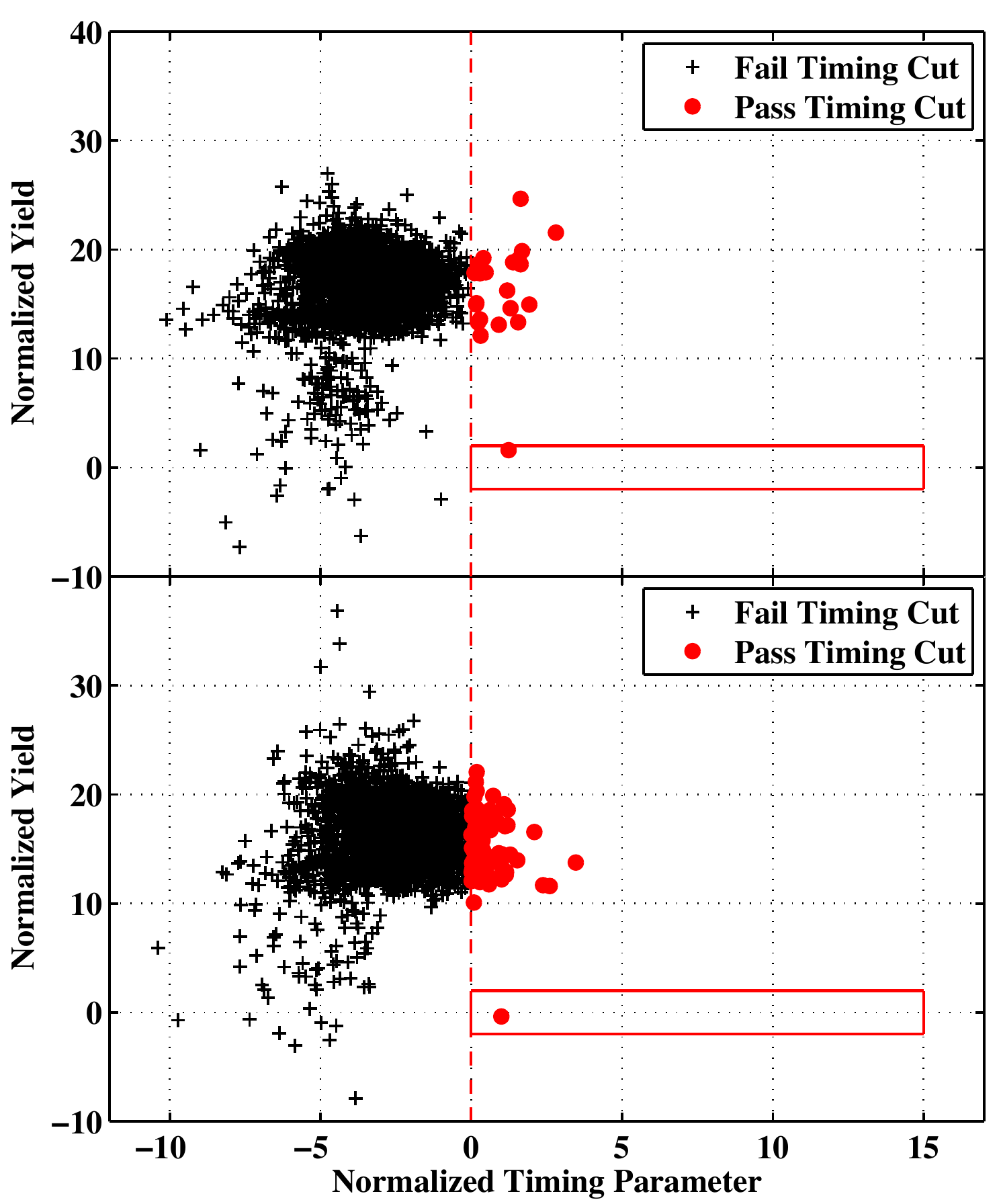}
\caption{Normalized ionization yield (number of standard deviations from mean of nuclear recoil band)
versus normalized timing parameter (timing relative to acceptance region) for events passing all cuts, excluding yield and timing.Ê 
The top (bottom) plot shows events for detector T1Z5(T3Z4).
Events that pass the phonon timing cut are shown with round markers.
The solid red box indicates the signal region for that detector.
The candidate events are the round markers inside the signal regions.  (Color online.)
 }
\label{fig:yield_timing}
\end{center}
\end{figure}

Upon unblinding, we observed two events in the WIMP acceptance region at recoil energies of
12.3\,keV and 15.5\,keV. These events are shown in
Figs.~\ref{fig:yield_energy} and \ref{fig:yield_timing}.

The candidate events occurred during periods of nearly ideal experimental performance, are
separated in time by several months, and occur in different towers. 
However, a detailed study revealed that an approximation made during the ionization pulse 
reconstruction degrades the timing-cut rejection of a small fraction of surface events with
ionization energy below $\sim$6\,keV. The candidate event in T3Z4 shows this effect. 
Such events are more prevalent in WIMP-search
data than in the data sets used to generate the pre-blinding estimate of misidentified surface events.
A refined calculation, which accounts for this reconstruction degradation, produced a revised surface-event estimate 
of $0.8 \pm 0.1 $(stat)$ \pm 0.2$(syst) events.  
The systematic uncertainty is dominated by our assumption that the pass-fail ratio for multiple scatter events is the same as that for single scatter events.
Based on this revised estimate, the probability to have observed two or more 
surface events in this exposure is 20\%; inclusion of the neutron background estimate increases this probability to 23\%. 
These expectations indicate that the results of this analysis cannot be interpreted as significant evidence for WIMP interactions, 
but we cannot reject either event as signal.

To quantify the proximity of these events to the surface-event rejection threshold, we varied the 
timing cut threshold of the analysis. Reducing the revised expected surface-event background to 0.4 
events would remove both candidates while reducing the WIMP exposure by 28\%.
No additional events would be added to the signal region until we increased the revised estimate of the expected 
surface-event background to 1.7 events.

\begin{figure}[t]
\begin{center}
\includegraphics[width=3.4in]{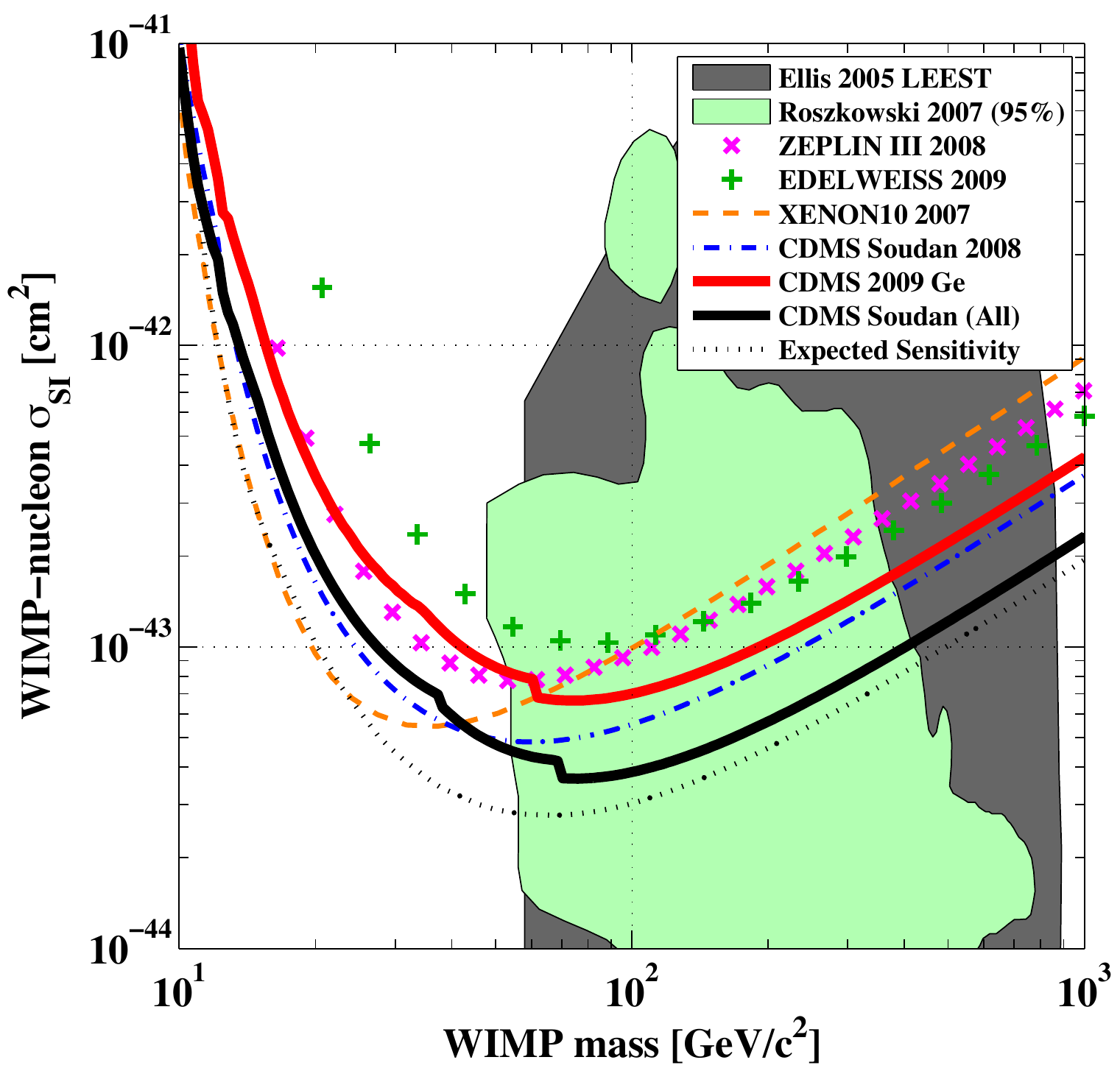} 
\caption{ 90\% C.L. upper limits on the WIMP-nucleon spin-independent cross section as a function of 
WIMP mass. The red (upper) solid line shows the limit obtained 
from the exposure analyzed in this work.  The solid black line shows the 
combined limit for the full data set recorded at Soudan. The dotted line indicates the expected sensitivity for 
this exposure based on our estimated background combined with the observed sensitivity of past Soudan data. 
Prior results from CDMS \cite{Ahmed:2008eu}, EDELWEISS II \cite{Armengaud:2009hc}, XENON10 \cite{Aprile:2008rc}, and ZEPLIN III \cite{Lebedenko:2008gb} 
are shown for comparison.  The shaded regions indicate allowed 
parameter space calculated from certain Minimal Supersymmetric Models \cite{Ellis:2005mb, Roszkowski:2007fd} (Color online.)}
\label{fig:silimit}
\end{center}
\end{figure}

We calculate an upper limit on the WIMP-nucleon elastic scattering cross-section based on standard galactic halo assumptions \cite{Lewin:1995rx} and in the presence of two events at the observed energies.  We use the Optimum Interval Method \cite{Yellin:2002xd} with no background subtraction.  The resulting limit shown in Fig. \ref{fig:silimit} has  a minimum cross section of $7.0 \times 10^{-44}$ cm$^{2}$ ($3.8 \times 10^{-44}$ cm$^{2}$ when combined with our previous results) for a 
WIMP of mass 70\,\gev\ .  The abrupt feature near the minimum of the new limit curve is a consequence of a threshold-crossing at which intervals containing one event enter into the optimum interval computation \cite{Yellin:2002xd}. An improved estimate of our detector masses was used for the exposure calculation of the present work; a similar correction (resulting in a $\sim$9\% decrease in exposure) was applied to our previous CDMS result \cite{Ahmed:2008eu} shown in Fig.~\ref{fig:silimit}.  
While this work represents a doubling of previously analyzed exposure, the observation of two events leaves the combined limit, shown in Fig.~ \ref{fig:silimit}, nearly unchanged below 60\,GeV/c$^2$ and allows for a modest strengthening in the limit above this mass.

\begin{figure}[t]
\begin{center}
\includegraphics[width=3.3in]{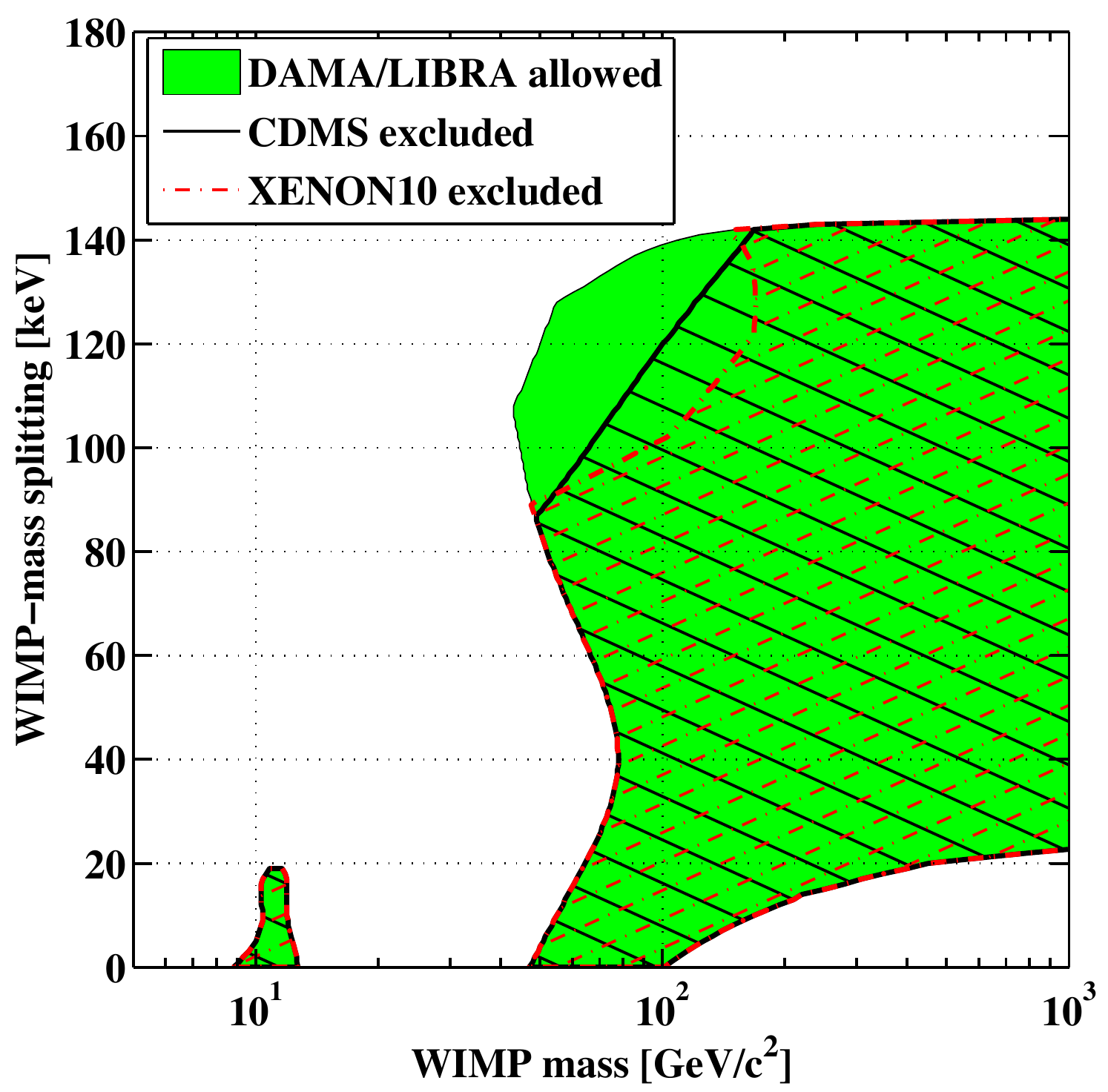} 
\caption{The shaded green region represents WIMP masses and mass splittings
for which there exists a cross section compatible with the DAMA/LIBRA \cite{Bernabei:2008yi} modulation spectrum at 90\% C.\,L.
under the inelastic dark matter interpretation \cite{TuckerSmith:2001hy}. 
Excluded regions for CDMS~II (solid-black hatched) and XENON10 \cite{Collaboration:2009xb} (red-dashed hatched) were 
calculated in this work using the Optimum Interval Method.  (Color online.)}
\label{fig:inelastic}
\end{center}
\end{figure}

We have also analyzed our data under the hypothesis of WIMP inelastic scattering \cite{TuckerSmith:2001hy}, which 
has been proposed to explain the DAMA/LIBRA data \cite{Bernabei:2008yi} . 
We computed DAMA/LIBRA regions allowed at the 90\% C.\,L. following the $\chi^{2}$ goodness-of-fit technique 
described in \cite{Savage:2008er}, without including channeling effects \cite{Bernabei:2007hw}.
Limits from our data and that of XENON10 \cite{Collaboration:2009xb} were computed using the 
Optimum Interval Method  \cite{Yellin:2002xd}. Regions excluded by CDMS and XENON10
were defined by demanding the 90\% C.\,L. upper limit to completely 
rule out the DAMA/LIBRA allowed cross section intervals for allowed WIMP 
masses and mass splittings. The results are shown in Fig.~\ref{fig:inelastic}.  
The CDMS data disfavor all but a narrow region of the parameter space allowed by DAMA/LIBRA that resides at a 
WIMP mass of $\sim$100\,\gev\ and mass splittings of 80--140 keV.

The data presented in this work constitute the final data runs of the CDMS II experiment and
double the analyzed exposure of CDMS II.   
We observed two candidate events.
These data, combined with our previous results, produce the strongest limit on spin-independent WIMP-induced 
nuclear scattering for WIMP masses above 42\,\gev\, ruling out new parameter space.

The CDMS collaboration gratefully acknowledges the contributions of numerous 
engineers and technicians; we would like to especially thank 
Jim Beaty, Bruce Hines, Larry Novak, 
Richard Schmitt and Astrid Tomada.  This work is supported in part by the 
National Science Foundation (Grant Nos.\ AST-9978911, PHY-0542066, 
PHY-0503729, PHY-0503629,  PHY-0503641, PHY-0504224, PHY-0705052, PHY-0801708, PHY-0801712, PHY-0802575 and PHY-0855525), by
the Department of Energy (Contracts DE-AC03-76SF00098, DE-FG02-91ER40688, 
DE-FG02-92ER40701, DE-FG03-90ER40569, and DE-FG03-91ER40618), by the Swiss National 
Foundation (SNF Grant No. 20-118119), and by NSERC Canada (Grant SAPIN 341314-07).

\bibliographystyle{apsrev}
\bibliography{arXiv_c58}

\end{document}